\documentclass[12pt]{iopart}
\usepackage{graphicx}
\usepackage{amssymb}

\newcommand{\Vext}{V_{\rm ext}}

\newcommand{\Vint}{V_{\rm int}}
\newcommand{\Vcm}{V_{\rm cm}}

\newcommand{\xcm}{x_{\rm cm}}

\newcommand{\vcm}{v_{\rm cm}}

\newcommand{\xcma}{\vec r_{\rm a,cm}}
\newcommand{\xcms}{\vec r_{\rm s,cm}}

\newcommand{\rcm}{\vec{r}_{\rm cm}}
\newcommand{\pcm}{\vec{p}_{\rm cm}}

\newcommand{\psicd}{\psi_{\rm cd}}
\newcommand{\Rcd}{R_{\rm cd}}
\newcommand{\Scd}{S_{\rm cd}}

\newcommand{\Mcm}{M_{\rm cm}}

\newcommand{\Qcm}{Q_{\rm cm}}
\newcommand{\sigmacm}{\sigma_{\rm cm}}

\newcommand{\im}{\mathop{\rm Im}}

\newcommand{\Err}{\mathop{\rm Err}}

\newcommand{\vy}{\vec{y}}

\begin{document}
\title{Conditions for the classicality of the center of mass of many-particle quantum states}

\author{Xavier Oriols}
\ead{xavier.oriols@uab.cat}
\address{Departament d'Enginyeria Electr\`onica, Universitat Aut\`onoma de Barcelona, 08193-Bellaterra (Barcelona), Spain}
\author{Albert Benseny}
\address{Quantum Systems Unit, Okinawa Institute of Science and Technology Graduate University, Onna, Okinawa 904-0495, Japan}

\begin{abstract}
We discuss the conditions for the classicality of quantum states with a very large number of identical particles. 
By treating the center of mass as a Bohmian particle, we show that it follows a classical trajectory when the distribution of the Bohmian positions in just one experiment is always equal to the marginal distribution of the quantum state in physical space.
This result can also be interpreted as a unique-experiment generalization of the well-known Ehrenfest theorem.
We also demonstrate that the classical trajectory of the center of mass is fully compatible with a conditional wave function solution of a classical non-linear Schr\"odinger equation.
Our work shows clear evidence for a quantum-classical inter-theory unification and opens new possibilities for practical quantum computations with decoherence. 
\end{abstract}

\maketitle

\section{Introduction}

Since the beginning of quantum theory a century ago, the study of the frontier between classical and quantum mechanics has been a constant topic of debate~\cite{landau,Ballentine90,herbert,Zurek03,Giulini96,diosi,schlosshauer14,Dieter}. Despite great efforts, the quantum-to-classical transition still remains blurry and certainly much more puzzling and intriguing than, for example, the frontier between classical mechanics and relativity. The relativistic equations of motion just tend to the classical ones when the velocities are much slower than the speed of light~\cite{herbert}.

The difficulties in finding a simple explanation for the classical-to-quantum transition have their roots in the so-called \emph{measurement problem} that requires getting rid of quantum superpositions~\cite{Dieter,maudlin,bohm66}.
Possible quantum states of a particle are represented by vectors in a Hilbert space.
Linear combinations of them, for example a superposition of macroscopically distinguishable state,  also correspond to valid states of the Hilbert space.
However, such superposition of states is not always compatible with measurements~\cite{bohm66,nikolic}.
The measurement problem can be formulated as the impossibility for a physical quantum theory (in empirical agreement with experiments) to satisfy  simultaneously the following three assumptions~\cite{maudlin}.
First, the wave function always evolves deterministically according to the linear and unitary Schr\"odinger equation.
Second, a measurement always find the physical system in a localized state, not in a superposition of macroscopically distinguishable states.
Third, the wave function is a complete description of a quantum system.
Different physical theories appear depending on which assumption is ignored~\cite{herbert}.

The first type of solutions argues that the unitary and linear evolution of the Schr\"odinger equation is not always valid.
For instance, in the instantaneous collapse theories~\cite{bassi13} (like the GRW interpretation~\cite{ghirardi86}), a new stochastic equation is used 
that breaks the superposition principle at a macroscopic level, while still keeping it at a microscopic one~\cite{bassi13}.
Another possibility is substituting the linear Schr\"odinger equation by a non-linear collapse law only when a measurement is performed~\cite{landau,bohr20}.
This is the well-known orthodox (or Copenhagen) solution, and most of the attempts to reach a quantum-to-classical transition have been developed under this last approach~\cite{Zurek03,Giulini96,diosi,schlosshauer14,Dieter,kim,Brukner,Yang13}.

A second type of solution ignores the assumption that a measurement always find the physical system in a localize state.
One then assumes that there are different worlds where different states of the superposition are found.
This is the many worlds solution~\cite{Everett57,wallace12,saunders}, in which the famous Schr\"odinger's cat
is found alive in one world and dead in another.
Explanations of the quantum-to-classical transition have also been attempted within this interpretation~\cite{saunders}.

There is a final kind of solutions that assumes that the wave function alone does not provide a complete description of the quantum state, i.e., additional elements (hidden variables) are needed.
The most spread of these approaches is Bohmian mechanics~\cite{bohm66,bohm52,bell66,Holland93,Oriols12,ABM_review,durr13}, where, in addition to the wave function, well-defined trajectories are needed to define a complete (Bohmian) quantum state. In a spatial superposition of two disjoint states in a single-particle system, only the one whose support contains the position of the particle becomes  relevant for the dynamics.   
Previous attempts to study the quantum-to-classical transition with Bohmian mechanics mainly focused on single-particle problems~\cite{durr13,sevensteps,Salvador13,Allori09}.
In this paper, we generalize such works by analyzing when the center of mass of a many-particle quantum system follows a classical trajectory.

The use of the center of mass for establishing the classicality of a quantum state has some promising advantages.
The first one is related to the description of the initial conditions.
Fixing the initial position and velocity of a classical particle seems unproblematic,
while it is forbidden for a quantum particle due to the uncertainty principle~\cite{landau,bohr20}.
The use of the center of mass relaxes this contradiction: it is reasonable to expect that two experiments with the same preparation for the wave function will give quite similar values for the initial position and velocity of the center of mass when a large number of particles is considered, although the microscopic distribution of all (Bohmian) particles will be quite different in each experiment.

The second advantage is that it provides a \emph{natural} coarse-grained definition of a classical trajectory that coexists with the underlying microscopic quantum reality.
One can reasonably expect that the Bohmian trajectory of the center of mass of a large number of particles can follow a classical trajectory, without implying that each individual particle becomes classical. 
Therefore, the use of the center of mass allows a definition of the quantum-to-classical transition, while keeping a pure quantum behavior for each individual particle.

This article is structured as follows.
We begin by studying the conditions under which the center of mass of a quantum state behaves classically.
We then present a type of wave functions that always fulfills these conditions, and show the equation that guides the wave function of the center of mass.
Next, we discuss examples of quantum states whose center of mass does not behave classically.
To finish, we summarize the main results, contextualize them within previous approaches and comment on further extensions of this work.

\section{Conditions for a classical center of mass}
\label{sec:conditions}

\subsection{Evolution of the center of mass in an ensemble of identical experiments}
\label{ensemble}

Throughout the article, we will consider a quantum system composed of $N$ particles of mass $m$ governed by the wave function $\Psi(\vec r_1,\ldots,\vec r_N,t)$ solution of the many-particle non-relativistic Schr\"odinger equation,
\begin{equation}
\rmi  \hbar \frac{\partial \Psi}{\partial t} = \left( -\frac{\hbar^2}{2m} \sum_{i = 1}^N \nabla^2_i + V\right) \Psi,
\label{mpscho}
\end{equation}
where $\vec r_i$ is the position of the $i$-th particle, $\nabla^2_i$ its associated Laplacian operator,
and the potential $V=V(\vec r_1,\ldots,\vec r_N,t)$ contains an external and an interparticle component,
\begin{equation}
V=\sum_{i=1}^{N} \Vext(\vec r_i)+\frac{1}{2}\sum_{i=1}^{N}\sum_{{f=1; i\neq f}}^{N}\Vint(\vec r_i-\vec r_f) .
\label{potential}
\end{equation}

In particular, we are interested in the evolution of one specific degree of freedom, the center of mass, defined as
\begin{equation}
\rcm = \frac{1}{N} \sum_{i=1}^{N}\vec r_i  .
\label{cm}
\end{equation}
Our aim in this paper is to analyze under which circumstances the observable associated to the operator $\rcm$ follows a classical trajectory in a unique experiment.

We first consider an ensemble of experiments realized with the same (prepared) wave function, 
whose average ensemble value of the center of mass is given by
\begin{equation}
\label{cmev}
\langle \rcm\rangle(t)=\int d^3\vec{r}_1 \ldots \int d^3\vec{r}_N   |\Psi(\vec r_1,\ldots,\vec r_N,t)|^2 \rcm.
\end{equation}
From Ehrenfest's theorem~\cite{ehrenfest27}, it is well-known that the time derivative of $\langle \rcm\rangle$ is
\begin{equation}
\label{eren1}
\frac{d \langle \rcm\rangle}{dt}=\frac{1}{N} \sum_{i=1}^{N} \frac{d \langle \vec r_i \rangle}{dt}=\frac{1}{N} \sum_{i=1}^{N} \langle \vec p_i \rangle=\langle \pcm \rangle .
\end{equation}
We can follow the same procedure for the time derivative of the momentum of the center of mass,
\begin{equation}
\label{eren2}
\frac{d \langle \pcm\rangle}{dt}=\frac{1}{N} \sum_{i=1}^{N} \frac{d \langle \vec p_i \rangle}{dt}= -\frac{1}{N} \sum_{i=1}^{N} \langle \nabla_i \Vext(\vec r_i) \rangle .
\end{equation}
When the spatial extent of the many-particle wave function is much smaller than the variation length-scale of the potential, we can assume $\langle \nabla_i \Vext(\vec r_i) \rangle = \nabla \Vext( \langle \rcm \rangle)$, and write
\begin{equation}
\label{eren3}
\frac{d^2 \langle \rcm\rangle}{dt^2}= - \nabla \Vext( \langle \rcm \rangle) .
\end{equation}
This classical behavior of the average $\langle \rcm\rangle$ is a very well-known result~\cite{landau,Ballentine90,ehrenfest27}.
The types of $\Vext$ that satisfy the condition $\langle \nabla_i \Vext(\vec r_i) \rangle = \nabla \Vext( \langle \rcm \rangle)$ will be further discussed later.

\subsection{Evolution of the center of mass in a unique experiment}
\label{evolution}

In order to satisfy our classical intuition, we need to certify that the observable associated to $\rcm$ follows a classical trajectory in each experiment (not in an average over several experiments).
This problem could be analyzed within the orthodox formalism~\cite{Zurek03,Giulini96,schlosshauer14,kim,Brukner,Yang13,lusanna}.
The typical approach would be to construct a reduced density matrix of the center of mass
by tracing out the rest of degrees of freedom interpreted as the environment.
The effect of decoherence, i.e. the entanglement between the environment and the system, then leads to a diagonal (or nearly diagonal) density matrix.
Finally, after invoking the collapse law, one obtains the observable result for the operator $\rcm$ by selecting one element of the diagonal at each measuring time.
In this work, however, we will approach the problem using Bohmian mechanics~\cite{bohm52,Holland93,Oriols12,ABM_review,durr13}.
This alternative formalism will allow us to reach the quantum-to-classical transitions without dealing with the reduced density matrix and without specifying the collapse law (this law is not needed in the Bohmian postulates~\cite{Holland93,Oriols12,durr13}).

As indicated in the introduction, in Bohmian mechanics, a quantum state is completely described by two elements:
the many-particle wave function $\Psi(\vec r_1,\ldots,\vec r_N,t)$ solution of the usual Schr\"odinger equation
and the trajectory $\{\vec r^j_i(t)\}$ of each $i=1 \ldots N$ particle.
Hereafter, each Bohmian quantum state will refer to a wave function and to a particular set of trajectories labeled by the superindex $j$ that correspond to a unique experiment. 
The velocity of each particle is given by
\begin{equation}
\label{velo}
\vec v^j_i(t)=\frac{d \vec r^j_i(t)}{dt}=\frac {\vec J_i(\vec r^j_1(t),\ldots,\vec r^j_N(t),t)}{|\Psi(\vec r_1(t),\ldots,\vec r_N(t),t)|^2} ,
\end{equation}
where $\vec J_i = \hbar \im(\Psi^* \nabla_i \Psi)/m$.
Thus, the configuration of particles reproduce all quantum features while evolving ``choreographed'' by the wave function~\cite{Oriols12,ABM_review,durr13,velocity,Marian16}. 

By construction, Bohmian predictions are as uncertain as the orthodox ones~\cite{FNL2016b}:
it is not possible to know the initial positions in a particular experiment (unless the wave function is a position eigenstate).
The best we can know about the particle positions in the $j$-experiment, $\{\vec r^j_i(t)\}$, is that they are found in locations where the wave function has a reasonable presence probability.
In particular, the set of positions in $M$ different experiments (prepared with the same wave function) are distributed according to
\begin{equation}
\label{QE}
|\Psi(\vec r_1,\ldots,\vec r_N,t)|^2 = \lim_{M \rightarrow \infty} \frac{1}{M} \sum_{j=1}^{M} \prod_{i=1}^N \delta\left(\vec r_i-\vec r^j_i(t)\right).
\end{equation}
If the set of $N$ positions follows this distribution at some time $t_0$, it is easy to demonstrate that \eref{QE} will also be satisfied at any other time $t$, provided that the many-particle wave function evolves according to \eref{mpscho} and that the particles moves according to \eref{velo}.
This property is known as equivariance~\cite{conditional} and it is key for the empirical equivalence between Bohmian mechanics and other quantum theories.
Equation \eref{QE} says that Born's law is always satisfied by counting particles~\cite{bohm52,Holland93,ABM_review,durr13} and that quantum results are unpredictable~\cite{FNL2016b}.
Several authors assume as a postulate of the Bohmian theory that the initial configuration of particles satisfies \eref{QE}, while others argue that it is just a consequence of being in a ``typical'' Universe~\cite{Callender,conditional}\footnote{In principle, one could postulate \eref{QE} (at some initial time) in the Bohmian theory in the same way that Born's law is a postulate in the orthodox theory.
However, some authors argue that this is not necessary~\cite{Callender}.
Probably the most accepted view against taking \eref{QE} as a postulate comes from the seminal work by D\"urr, Goldstein, and Zangh\`i~\cite{conditional}, where the equivariance in any system is discussed from the initial configurations of (Bohmian) particles in the Universe.
Using Bohmian mechanics to describe the wave function of the whole Universe,
then the wave function associated to any (sub)system is an effective (conditional) wave function of the universal one.
Using typicality arguments, D\"urr \etal showed that the overwhelming majority of possible selections of initial positions of particles in the Universe will satisfy the condition \eref{QE} in a subsystem~\cite{conditional}.
Other authors~\cite{valentini} have attempted to dismiss \eref{QE} as a postulate by showing that any initial configuration of Bohmian particles will relax, after some time, to a distribution very close to \eref{QE} for a subsystem. }.

After selecting one initial positions of the particles from \eref{QE} in a unique $j$-experiment, we can then define the trajectory for the center of mass of the Bohmian quantum state associated to such $j$-experiment as
\begin{equation}
\rcm^j(t) = \frac{1}{N} \sum_{i=1}^{N}\vec r^j_i(t) .
\label{cmue}
\end{equation}
As discussed above, in general $\rcm^j(t)\neq \rcm^{h}(t)$ for any two different experiments $j$ and $h$, because the Bohmian positions have an intrinsic uncertainty coming from \eref{QE}.

\subsection{Classical center of mass in a unique experiment}
\label{conditions}

A classical trajectory for the center of mass $\rcm^j(t)$ of a quantum state in a unique experiment is obtained  when the following two conditions are satisfied:
\begin{itemize}
\item \textbf{Condition 1} ---
For the overwhelming majority of experiments associated to the same wave function,
the same trajectory for the center of mass is obtained.
That is to say, for (almost) any two different experiments $j$ and $h$ we obtain $\rcm^j (t) = \rcm^{h}(t)$.

\item \textbf{Condition 2} ---
The spatial extent of the (many-particle) wave function 
in each direction
is much smaller than the variation length-scale of the external potential $\Vext$.
\end{itemize}

According to condition 1, since $\rcm^{j}(t)=\rcm^{j0}(t)$ for all $M$ experiments, the empirical evaluation of $\langle \rcm\rangle$ will be equal to the trajectory of the center of mass $\rcm^{j0}(t)$ in a unique experiment:
\begin{equation}
\label{condition2}
\langle \rcm\rangle(t)=\lim_{M \rightarrow \infty} \frac {1}{M} \sum_{j=1}^M \rcm^j(t) = \rcm^{j_0}(t) .
\end{equation}
Moreover, we notice that $\rcm^{j}(t)$ in such quantum state has the same well-defined initial conditions (position and velocity) as in the overwhelming majority of experiments. While condition 1 might seem very restrictive, we will show in what follows that quantum states that satisfy it are more natural than expected when the number of particles is very large.

A better understanding of condition 2 can be found from a Taylor expansion of the external potential $\Vext(\vec r_i)$ in \eref{eren2}.
One can easily realize that the condition
$\langle \nabla \Vext(\vec r_i) \rangle = \nabla \Vext( \langle \vec r_i \rangle)$
is directly satisfied by constant, linear or quadratic potentials.
Where $\Vext$ can be approximated by potentials with such dependence requires a discussion on its physical meaning.
$\Vext(\vec r_i)$ in \eref{potential} describes the interaction of particle $i$ with some distant ``source'' particles located elsewhere.
Moreover, the fact that this potential is felt identically by all $N$ system particles (i.e.  $\Vext(\vec r_i)$ is a single particle potential) is due to the large distance between our system and the potential sources.
We can then assume, that $\Vext$ is generated by some kind of long-range force, such as electromagnetic or gravitational ones.
Such external long-range potentials will usually have a small spatial variation along the support of $\Psi(\vec r_1,\ldots,\vec r_N,t)$ and a linear or quadratic approximation for $\Vext$ would seem enough in most macroscopic scenarios.
In any case, scenarios where higher orders of the series expansion of $\Vext$ are relevant are possible in the laboratory. Then, 
if condition 1 is applicable, it will guarantee a unique trajectory $\langle \rcm\rangle(t)=\rcm^{j_0}(t)$ in all experiments with well-defined initial conditions, however its acceleration will not only be given by the gradient of $\Vext$, but it will also depend on the wave function.

\section{Quantum states with a classical center of mass}
\label{sec:fullofparticles}

\subsection{Quantum state full of identical particles}
\label{example2}

We define here a type of quantum state with a very large number of indistinguishable particles (either fermions or bosons) that we name \emph{quantum state full of identical particles}.
We will show that the center of mass of these states always follows a classical trajectory. 
Our definition will revolve around the concept of marginal probability distribution, i.e. the spatial distribution for the $i$th particle independently of the position of the rest of the particles, i.e.,
\begin{equation}
D(\vec r_i,t)
= \int\ldots\int |\Psi(\vec r_1,\ldots,\vec r_N,t)|^2 \prod_{{f=1;  f\neq i}}^N d^3\vec{r}_f .
\label{margdef1}
\end{equation}
Empirically, this distribution can be calculated from a very large number $M$ of experiments as
\begin{equation}
D(\vec r_i,t) =\lim_{M \rightarrow \infty} \frac{1}{M} \sum_{j=1}^M \delta( \vec r_i-\vec r_i^j(t)) .
\label{margdef2}
\end{equation}
Since our definition of a quantum state full of identical particles always involves indistinguishable particles, the subindex $i$ is superfluous, and all particles will have the same marginal distribution.
We notice that, while all Bohmian particles $\vec r_i(t)$ are ontologically distinguishable (through the index $i$), the Bohmian dynamical laws, Eqs. \eref{mpscho} and \eref{velo}, ensure that they are empirically indistinguishable\footnote{
The empirical indistinguishability of the Bohmian trajectories means that the $\vec r_2$-observable computed from $\vec r^j_2(t)$ is identical to the $\vec r_1$-observable computed from $\vec r^j_1(t)$. This property can be easily understood from the symmetry of the wave function, see also Refs.~\cite{Holland93,durr13,identical}.
Consider a set of trajectories $\{\vec r^j_1(t),\vec r^j_2(t),\ldots,\vec r^j_N(t)\}$ assigned to an experiment $j$.
We construct another set of trajectories $\{\vec r^h_1(t),\vec r^h_2(t),\ldots,\vec r^h_N(t)\}$ whose initials conditions are $\vec r^h_1(0)=\vec r^j_2(0)$ and $\vec r^h_2(0)=\vec r^j_1(0)$, while $\vec r^h_i(0)=\vec r^j_i(0)$ for $i=3,\ldots,N$.
Due to the symmetry of the wave function (and of the velocity \eref{velo}), $\vec r^h_1(t)=\vec r^j_2(t)$ and $\vec r^h_2(t)=\vec r^j_1(t)$ (the rest of trajectories are identical in $j$ and $h$).
Any observable related to $\vec r_1$ (or $\vec r_2$) is evaluated over an ensemble of different experiments.
For each $j$-element of the ensemble, we can construct its corresponding $h$-set of trajectories and evaluate the $\vec r_2$-observable using $\vec r_2^h(t)$ instead of $\vec r_2^j(t)$.
By construction, since $\vec r^h_2(t)=\vec r^j_1(t)$, the $\vec r_2$-observable is identical to the $\vec r_1$-observable.}.

We define a quantum state full of identical particles as a state whose distribution of the positions of the $N$ particles in just one experiment is always equal to the marginal distribution of a unique variable obtained from averaging over different experiments,
\begin{equation}
D(\vec r,t) = \lim_{N \rightarrow \infty} \frac{1}{N} \sum_{i=1}^N \delta(\vec r-\vec r_i^{j_0}(t))  = \lim_{M \rightarrow \infty} \frac{1}{M} \sum_{j=1}^M \delta( \vec r-\vec r_i^j(t)).
 \label{mar1}
\end{equation}
For the practical application of this definition in systems with a finite (but very large) number of particles, one can impose that the condition in \eref{mar1} has to be satisfied for the overwhelming majority of experiments, see \ref{app:error}.

The selection of the initial position of the particles, $\vec r_1^{j_0}(0),\vec r_2^{j_0}(0)\ldots \vec r_N^{j_0}(0)$, in a single experiment (labeled here $j_0$) can be done from \eref{QE}.
One would start by first selecting $\vec r_1^{j_0}(0)$ (independently of the rest of positions).
Then, selecting $\vec r_2^{j_0}(0)$ conditioned to the fact the $\vec r_1^{j_0}(0)$ is already selected.
This procedure is repeated until the last position is selected, $\vec r_N^{j_0}(t)$, conditioned to all previous selected positions.
The probability distribution for selecting the trajectory $\vec r_i^{j_0}(0)$ , when the previous positions $\vec r_1^j(0),\ldots ,\vec r_{i-1}^{j_0}(0)$ are already selected, can be defined from a combination of conditional and marginal probabilities as:
\begin{equation}
D^{j_0,i}(\vec r_i,0) = \frac{\bar D^i(\vec r_1^{j_0}(0),\ldots ,\vec r_{i-1}^{j_0}(0),\vec r_{i},0)}{\int \bar D^i(\vec r_1^{j_0}(0),\ldots ,\vec r_{i-1}^{j_0}(0),\vec r_{i},0) d\vec r_{i}}
 \label{marcon1}
\end{equation}
with
\begin{equation} \fl
\bar D^i(\vec r_1,\ldots ,\vec r_{i},0) = \int\ldots\int |\Psi(\vec r_1,\ldots ,\vec r_{i},\ldots,\vec r_N,0)|^2 d^3\vec{r}_{i+1}\ldots d^3\vec{r}_N
 \label{marcon2}
\end{equation}
By construction, the probability distribution function in \eref{mar1} has a total probability equal to unity. On the contrary, a normalization constant is explicitly included in the definition of \eref{marcon1} to ensure that it is a probability distribution function properly normalized to unity. In particular, for any $j_0$-experiment, we get $D^{j_0,1}(\vec r_1,0) \equiv D(\vec r_1,0)$ and $D^{j,N}(\vec r_N,0) \equiv |\Psi(\vec r_1^{j_0}(0),\ldots ,\vec r_{i}^{j_0}(0),\ldots,\vec r_{N-1}^{j_0}(0),\vec r_N,t)|^2$. Therefore, a quantum state full of identical particles can be alternatively defined as the wave function satisfying that the global distribution of the $i=1,\ldots,N$ particles in a unique $j_0$-experiment constructed from \eref{marcon1} and \eref{marcon2}, is equal to $D(\vec r,0)$ in \eref{margdef1} for the overwhelming majority of experiments. A trivial example of a quantum state full of identical particles is the one where the corresponding distribution for selecting the $i=1,\ldots,N$ particles in the overwhelming majority of experiments satisfies $D^{j_0,i}(\vec r_i,0) = D(\vec r_i,0)$.

The equivalence between both expressions in \eref{mar1} implies the equivalence between two sets of positions:
first, the positions of particle $i_0$ in $M$ different experiments,
$\{\vec r_{i_0}^j(t)\}$ for $j=1,\ldots,M$,
and, second, the positions of the $N$ particles in the same $j_0$-experiment,
$\{\vec r_i^{j_0}(t)\}$ for $i=1,\ldots,N$.
Because of this equivalence, a position in the first set, say $\vec r_i^{j_0}(t)$, is equal to another position in the second set, $\vec r_{i_0}^j(t)$. Any position of one set has another identical position in the other set.  
Therefore, since the exchange of positions of identical particles does not exchange their velocity~\cite{identical}, we obtain that $\vec v^{j_0}_i=\vec v_{i_0}^j$, which implies that $\vec r_i^{j_0}(t) =\vec r_{i_0}^j(t)$ at any time.
Therefore, we conclude that if \eref{mar1} is satisfied at a particular time, such as $t=0$, then the quantum state will be full of identical particles at any other time.

At this point, using \eref{mar1} for any time $t$, we can certify that the trajectory of the center of mass of a quantum state full of identical particles satisfies,
\begin{eqnarray} \fl
\rcm^{j_0}(t)
=
\lim_{N\to\infty}
\frac{1}{N} \sum_{i=1}^{N} \vec r^{j_0}_i(t)
=
\lim_{M\to\infty}
\frac{1}{M} \sum_{j=1}^{M} \vec r^j_{i_0}(t) 
\nonumber \\
=
\lim_{N,M\to\infty}
\frac {1} {N} \sum_{i=1}^{N} \frac{1}{M} \sum_{j=1}^{M} \vec r^j_i(t) 
=
\lim_{N,M\to\infty}
\frac{1}{M} \sum_{j=1}^{M}  \frac {1} {N} \sum_{i=1}^{N} \vec r^j_i(t) 
\nonumber \\
=
\lim_{M\to\infty}
\frac{1}{M} \sum_{j=1}^{M} \rcm^j = \langle \rcm \rangle(t), 
\label{qsfip}
\end{eqnarray}
where we have used that
\begin{equation}
\rcm^{j_0}(t)=\int \vec r\;  D(\vec r,t) d\vec r
\label{cmfromd}
\end{equation}
with $D(\vec r,t)$ given by any of the two expressions in \eref{mar1}. In summary, a quantum state full of identical particles satisfies condition 1, and, if condition 2 also holds, its center of mass will be a classical trajectory.

The arguments we have presented here is for a system of indistinguishable particles.
For a macroscopic object composed of several types of particles, we can apply the same reasoning and obtain a classical center of mass for each type of particle subsystem, such that the global center of mass is also classical.

\subsection{Example 1: Many-particle quantum state with a unique single-particle wave function}
\label{example1}

Here we show the simplest example of a quantum state full of identical particles. 
We consider a $N$-particle wave function given by
\begin{equation}
\Psi(\vec r_{1},\ldots, \vec r_{N},t) =\prod_{j=1}^{N} \psi (\vec r_j,t) .
\label{example11}
\end{equation}
It corresponds, for example, to a system of non-interacting bosons, all with the same single-particle wave function $\psi(\vec r,t)$ solution of a single-particle Schr\"odinger equation under the external potential $\Vext(\vec r)$.

The quantum state in the $j$-experiment is completed with the set of trajectories $\{\vec r_i^j(t)\}$ for $i=1,\ldots,N$ selected according to $|\Psi|^2$.
Since \eref{example11} corresponds to a separable system, each position $\vec r_i^j(0)$ has to be selected according to its own probability distribution in \eref{marcon1} and \eref{marcon2} with $D^{j_0,i}(\vec r_i,0) =|\psi(\vec r_i,0)|^2$. The marginal distribution in \eref{margdef1}  satisfies $D(\vec r_i,0)=|\psi(\vec r_i,0)|^2$, which is exactly the same distribution mentioned above for selecting the particles. Therefore, this quantum state trivially satisfies \eref{mar1} when $N \rightarrow \infty$, i.e. $D^{j_0,i}(\vec r,0) = D(\vec r,0)$. As a result, the (Bohmian) trajectory of the center of mass will follow a classical trajectory when condition 2 about $\Vext$ is also satisfied. 

\subsubsection*{Numerical example}

\begin{figure}
\centerline{\includegraphics[width=0.6\columnwidth]{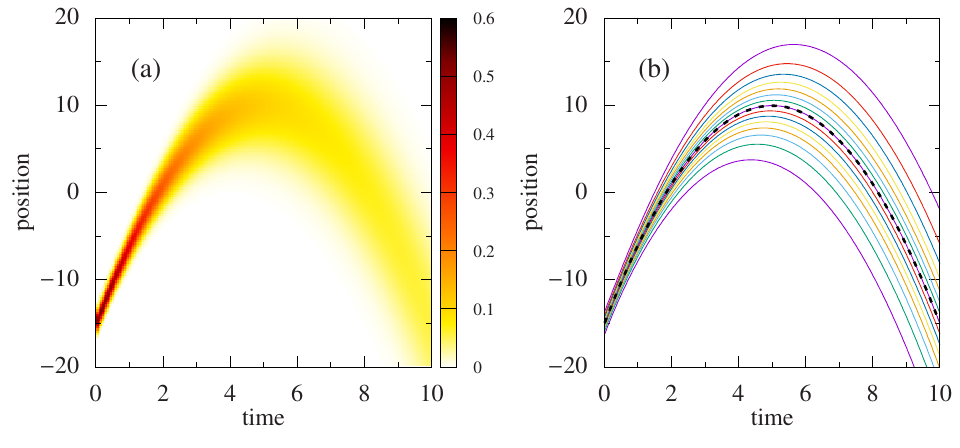}}
\caption{
(a)
Evolution of a quantum wave packet with a potential $\Vext(x)=2 x$.
The initial wave function is a Gaussian wave packet of width $\sigma = 1$, centered around $x_0=-15$, and an initial positive velocity $k_0 = 10$.
(b)
Quantum trajectories corresponding to the dynamics in (a); with the average shown as a the dashed black line.
Units are $m=\hbar=1$.
}
\label{fig:Q1}
\end{figure}

\begin{figure}
\centerline{\includegraphics[width=0.6\columnwidth]{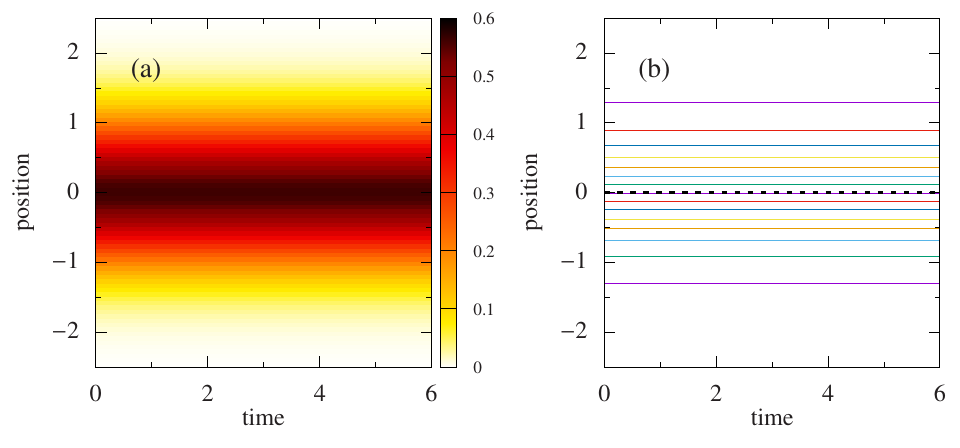}}
\caption{
Same as \fref{fig:Q1} but for the evolution of an initial Gaussian wave packet with $x_0=0$, $\sigma=1$ and $k_0=0$, in a potential $\Vext(x)=x^2/2$.
}
\label{fig:Q2}
\end{figure}

For simplicity, we consider a 1D physical space to numerically test the properties of the above state.
As the initial single-particle wave function we select a wave packet of the form
\begin{equation}
\psi(x,0)= \frac{1}{\sqrt{\sigma\sqrt{\pi}}} \exp\left( -\frac{(x-x_0)^2}{2 \sigma^2} \right) \exp(i k_0 x) ,
\label{eq:wf}
\end{equation}
with $\sigma$ the dispersion of the wave-packet, $x_0$ the initial position and $k_0$ the initial momentum.
Then, since the particles are independently selected, the central limit theorem~\cite{central} ensures that the center of mass of the quantum state will be normally distributed with a dispersion $\sigmacm=\sigma/\sqrt{N} \rightarrow 0$, confirming that the center of mass has the same well-defined position in all experiments (see \ref{app:error}). 

In the first example in \fref{fig:Q1} we use a linear potential $\Vext(x) = 2 x$ emulating a particle in free fall under a gravity force. The quantum wave packet increases its width over time and its center follows a typical parabolic movement. The second example in \fref{fig:Q2} corresponds to a harmonic potential $\Vext(x)=x^2/2$. In this case, because the wave function corresponds to the ground state of the quantum harmonic oscillator, it does not show any dynamics and the trajectories remain static at their initial positions. In any case, the center of mass (dashed black line in \fref{fig:Q2}) corresponds to the classical trajectory at the position of the minimum of the harmonic potential with zero velocity.

Now, we confirm the classicality of the center of mass of a quantum state defined by \eref{example11} using simpler arguments. Since there is no correlation between different trajectories $x_i^j(t)$, the Bohmian trajectories plotted in figures~\ref{fig:Q1} and \ref{fig:Q2} can be interpreted in two different ways.
The first interpretation is the one explained above where they correspond to different $i=1,\ldots,N$ trajectories in the same experiment described by the many-particle wave function given by \eref{example11}.
In this case, the average value of the trajectories (dashed black lines in figures~\ref{fig:Q1}(b) and \ref{fig:Q2}(b)) is understood as the trajectory for the center of mass in that particular experiment.
The second interpretation is that the trajectories correspond to different experiments of a single particle system defined by the wave function $\psi(x,t)$. In this interpretation, $\langle \xcm \rangle$ corresponds to a classical trajectory (for large enough $N$ and $\Vext$ satisfying condition 2), as shown by Ehrenfest's theorem~\cite{ehrenfest27} discussed in \sref{ensemble}.
Since the trajectories in both interpretations are mathematically identical, we  conclude that the (Bohmian) trajectory of the center of mass in a unique experiment follows a classical trajectory $\xcm^j(t)=\langle \xcm \rangle$, as anticipated in the discussion above on how these quantum states satisfy the condition in \eref{mar1} , i.e. $D^{j,i}(x,0) = D(x,0)$.

\subsection{Example 2: Many-particle quantum state with exchange and inter-particle interactions}
\label{examfop}

In the following we consider a more general example of quantum state full of identical particles with exchange and inter-particle interactions. We consider here a quantum wave function $\Psi$ which, at time $t=0$, is build from permutations of $N$ single-particle wave functions, $\psi _{i}(\vec r,0)$. We define $\Psi(\vec r_{1},\ldots, \vec r_{N},0)$ as
\begin{equation}
\Psi(\vec r_{1},\ldots, \vec r_{N},0) = \sum_{\vec p \in S_N}\prod_{i=1}^{N} \psi _{p_i}(\vec r_i,0)  s_{\vec p}, 
\label{exch0}
\end{equation}
where $\vec p=\{p_1,p_2,\ldots,p_N\}$ is an element of the set $S_N$ of $N!$ permutations of $N$ elements. The term  $s_{\vec p} = \pm 1$ is the sign of the permutation for fermions, while $s_{\vec p}=1$ for bosons. A global normalization constant has been omitted because it will be irrelevant. In particular, we consider that the single-particle wave functions $\psi _{i}(\vec r,0)$ and $\psi _{j}(\vec r,0)$ are either identical or without spatial overlapping. For any $\vec r$ and $\psi _{f}(\vec r,0)$, we have:
\begin{eqnarray}
\label{defexch1}
\psi _{f}(\vec r,0)=\psi _{i}(\vec r,0)				\qquad &\forall f \in N_i,   \nonumber\\
\psi _{f}(\vec r,0) \psi _{i}(\vec r,0) \simeq 0	\qquad &\forall f \notin N_i,
\label{defexch2}
\end{eqnarray}
where $N_i$ is the subset of wave functions identical to $\psi _{i}(\vec r,0)$. We now check if the quantum state defined by Eqs. \eref{exch0} and \eref{defexch2} is a quantum state full of identical particles. The initial modulus squared of the wave function in \eref{exch0} can be written as
\begin{equation}
|\Psi|^2= \sum_{\vec p,\vec p' \in S_N}\prod_{i=1}^{N} \psi_{p_i}(\vec r_i,0) \psi^*_{p'_i}(\vec r_i,0) s_{\vec p} s_{\vec p'},
\label{modulexch}
\end{equation}
and the marginal distribution for each particle is then given from \eref{margdef1} as 
\begin{equation}
D(\vec r,0)= \sum_{\vec p,\vec p' \in S_N} \psi_{p_1}(\vec r,0) \psi^*_{p'_1}(\vec r,0) \prod_{i=2}^{N} d_{p_i,p'_i} s_{\vec p} s_{\vec p'}, 
\label{exch2}
\end{equation}
with the matrix element $d_{i,f}$ defined as 
\begin{equation}
d_{i,f} = \int \psi_{i}(\vec r,0) \psi^*_{f}(\vec r,0) d^3\vec r .
\label{exch3}
\end{equation}
Because of \eref{defexch2}, $d_{i,f} = 1$ for all $f \in N_i$ and $d_{i,f} \simeq 0$ for $f \notin N_i$. Then, only the summands in \eref{exch2} with all the terms $d_{i,f}=1$ are different from zero, and we can rewrite $D(\vec r,0)$ as
\begin{equation}
D(\vec r,0)= \alpha \left( \sum_{i=1}^N |\psi_{i}(\vec r,0)|^2 \right).
\label{exch22}
\end{equation}
where $\alpha$ is the product of the number of permutations of each $N_i$ to provide a properly normalized distribution in \eref{mar1}.

On the other hand, the selection of the $N$ positions in a unique experiment $\{\vec r_i^j(0)\}$ has to satisfy \eref{QE}.
The selection of the first particle $\vec r_1^j(0)$ (independently on all other particles) is given by \eref{exch22}.
To select the second particle $\vec r_2^j(0)$, one needs to take into account the already selected $\vec r_1^j(0)$.
In general, according to the definitions \eref{marcon1} and \eref{marcon2} and using \eref{modulexch}, \eref{exch2} and \eref{exch3}, the selection of the position $\vec r_{m}^j(0)$ as a function of the previous $m-1$ positions $\vec r_{1}^j(0),\ldots,\vec r_{m-1}^j(0)$ is given by the distribution
\begin{equation} \fl
D^{j,m}(\vec r,0) = \sum_{\vec p,\vec p' \in S_N} \left( \prod_{k=1}^{m-1}w^j_{k,p_k,p'_k}\right) \psi_{p_m}(\vec r,0) \psi^*_{p'_m}(\vec r,0)
\left( \prod_{i=m+1}^{N} d_{p_i,p'_i}\right)  s_{\vec p} s_{\vec p'} , 
\label{exch4}
\end{equation}
with the matrix element $w^j_{k,p_k,p'_k}$ defined as
\begin{equation}
w^j_{k,p_k,p'_k} =\psi_{p_k}(\vec r_k^j(0),0) \psi^*_{p'_k}(\vec r_k^j(0),0) .
\label{exch5}
\end{equation}
For each position $\vec r^j_k(0)$, because of \eref{defexch2}, there is a $N_i$ set of wave functions whose value is $w^j_{k,i,f}=|\psi _{i}(\vec r^j_k(0),0)|^2$ for any $f \in N_i$, and $w^j_{k,i,f} \simeq 0$ for any $f \notin N_i$.
Again, we can assume that only the summands with the products $w^j_{k,i,f}=|\psi _{i}(\vec r^j_k(0),0)|^2$ and $d_{i,f}=1$ will remain different from zero in \eref{exch4} giving $\psi_{i}(\vec r,0) \psi^*_{f}(\vec r,0)=|\psi_{i}(\vec r,0)|^2$. We can then rewrite $D^{j,m}(\vec r,0)$ as
\begin{eqnarray}
D^{j,m}(\vec r,0) = \beta_m \left( \sum_{i=1}^N |\psi_{i}(\vec r,0)|^2 \right)
\label{exch6}
\\
\beta_m = \alpha \sum_{\vec p \in S_{M-1}} \prod_{k=1}^{m-1} |\psi _{p_k}(\vec r^j_k(0),0)|^2 .
\end{eqnarray}
Again, the parameter $\beta_m$ is irrelevant because the selection of the particles can be done through an expression of $D^{j,m}(\vec r,0)$ properly normalized to unity, where only the dependence on $\vec r$ matters. 

In summary, for the quantum state defined by Eqs.~\eref{exch0} and \eref{defexch2} plus a set of trajectories $\{\vec r_i^j(0)\}$, we conclude that the (normalized versions of the) distributions $D(\vec r,0)$ in \eref{exch22} and $D^{j,m}(\vec r,0)$ in \eref{exch6} for any $m$ are identical. Therefore we are dealing with a quantum state full of identical particles whose center of mass follows a classical trajectory.   

As we have demonstrated in \sref{example2}, whether $\Psi(\vec r_{1},\ldots, \vec r_{N},t)$ fulfills the condition in \eref{mar1} or not has to be tested in a unique time. Since we have shown that \eref{exch0} is a quantum state full of identical particles at $t=0$,  we conclude that any quantum state with the wave function $\Psi(\vec r_{1},\ldots, \vec r_{N},t)$ solution of the many-particle Schr\"odinger equation in \eref{mpscho}, with or without external $\Vext$ or inter-particle $\Vint$ potentials, and with the initial state defined by Eqs.~\eref{exch0} and \eref{defexch2} is a quantum state full of identical particles when $N \rightarrow \infty$.  

\subsubsection*{Numerical example}

\begin{figure}
\centerline{\includegraphics[width=0.6\columnwidth]{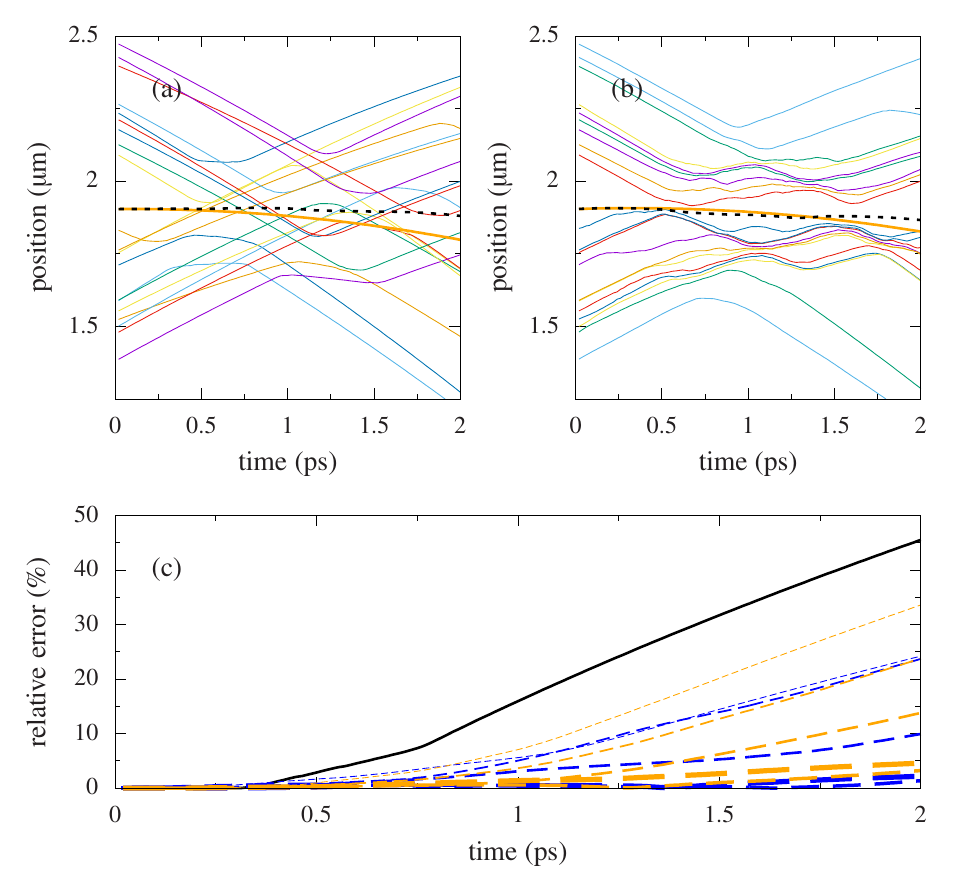}}
\caption{
(a) Simulation with $N=20$ distinguishable particles:
particle trajectories (thin lines),
quantum center of mass trajectory (dashed black line),
classical center of mass trajectory (solid orange line).
(b) Same as (a) but for indistinguishable particles.
(c) Relative error between the classical and quantum center of mass trajectories for
1 particle (black solid line) or $N$ distinguishable (light orange) or indistinguishable (dark blue) particles.
From thin to thick lines: $N=4$, 8, 12, 16, and 20 particles.
}
\label{fig:one}
\end{figure}

In what follows we investigate numerically this system. We will show that the center of mass of the quantum state effectively tends to a classical result even for a quite small number of particles.  
The evolution of the initial wave function in \eref{exch0} in the limit of $N \rightarrow \infty$ is numerically intractable. 
We will consider here a finite number of non-interacting bosons in a 1D space and test if the center of mass tends to a classical trajectory when $N$ increases.
Each single-particle wave function $\psi_i(x_{i},t)$ is a solution of a single-particle Schr\"odinger equation under the potential $\Vext$.
Therefore, the bosonic many-particle wave function can be written at any time $t$  as
\begin{equation}
\Psi(x_{1},\ldots, x_{N},t) = \sum_{\vec p \in S_N}\prod_{i=1}^{N} \psi _{p_i}(x_i,t) , 
\label{exch}
\end{equation}    
For comparison, we also consider the same state in \eref{exch}, but without exchange interaction
\begin{equation}
\Psi(x_{1},\ldots, x_{N},t) = \prod_{i=1}^{N} \psi_i(x_{i},t).
\label{no-ex}
\end{equation}
In particular, we will consider each of the $\psi_i$ in Eqs.~\eref{exch} and \eref{no-ex} as a sum of two initially separated Gaussian wave packets, but with opposite central momenta to ensure that they impinge at a later time
\begin{equation} \fl
\psi _i (x_j,0 ) = \frac{\exp {\left( {ik_{iL} x_j } \right)}}{{2\left( {\pi \sigma^2 } \right)^{1/4} }} \exp{\left( {- \frac{{\left( {x_j - x_{iL} } \right)^2 }}{{2\sigma^2 }}} \right)}
+\frac{\exp {\left( {ik_{iR} x_j } \right)}}{{2\left( {\pi \sigma^2 } \right)^{1/4} }}  \exp{\left( {- \frac{{\left( {x_j - x_{iR} } \right)^2 }}{{2\sigma^2 }}} \right)},
\label{gausiana}
\end{equation}
The $x_{iL}$ and $x_{iR}$ are the centers of two (non-overlapping) Gaussian wave packets, with respective momenta $k_{iL}$ and $k_{iR}$, and spatial dispersion $\sigma=15$ nm.
Each of the wave functions have different random values for $x_{iL}$, $x_{iR}$, $k_{iL}$, and $k_{iR}$.
These wave functions are evolved using Schr\"odinger equation with an external potential $\Vext$ implying a constant electric field of $3.3 \times 10^5$ V/m.

We show in \fref{fig:one}(a,b) for the cases with and without exchange interaction, the evolution of the quantum trajectories (thin lines). We plot their quantum center of mass (dashed black line) computed from \eref{cmue} for $N=20$.
We also plot the classical center of mass (solid orange line), computed from a Newtonian trajectory with the same initial position and velocity as the previous quantum center of mass. We notice that the Bohmian trajectories for states with exchange interaction do not cross in the physical space. This is a well-know property~\cite{identical} that obviously remains valid even if the center of mass becomes classical.  

Moreover, in \fref{fig:one}(c) we show the difference between the quantum and classical centers of mass for different values of $N$, with and without exchange interaction (see \ref{app:error} for a discussion of the error of a quantum state full of identical particles when a large, but finite, number of particles is considered).
We see that the quantum center of mass $\xcm(t)$ becomes more and more classical as $N$ grows, and the indistinguishable case reduces the quantum non-classical effects faster than the case without exchange interaction.
These results can be interpreted in a simple way:
a unique experiment with $N$ distinguishable particles represents effectively only one experiment,
while a unique experiment with $N$ indistinguishable particles represents, in fact, $N!$ different experiments, each one with the initial (Bohmian) positions interchanged. This explains why the latter center of mass become more similar to that given by the Ehrenfest theorem which involves an infinite number of experiments.

\subsection{Wave equation for the center of mass}
\label{wavequation}

While the description of a classical state requires only a trajectory, a complete Bohmian quantum state requires a wave function plus trajectories.
Moreover, because of its exponential complexity, solutions to the Schr\"odinger equation in the whole many-particle configuration space are not accessible.
However, an equation describing the evolution of a wave function associated to the center of mass of a quantum state full of identical particles will help to certify that a classical center of mass behavior is fully compatible with a \emph{pure} quantum state.
In addition, such an equation will provide an accessible numerical framework to analyze practical quantum system under decoherence.
One route towards this equation could be obtained from the reduced density matrix of the center of mass, and assuming some kind of collapse.
Alternatively, as mentioned along the paper, we will follow a Bohmian procedure which allows the construction of such a wave equation for the center of mass through the use of the (Bohmian) conditional wave function~\cite{conditional,Oriols07,norsen}.

To simplify the derivations, in the following we restrict ourselves to a 1D physical space.
We define the center of mass of our $N$-particle state, $\xcm$, and a set of relative coordinates, $\vy=\{y_2,\ldots,y_N\}$, as
\begin{eqnarray}
\xcm = \frac{1}{N} \sum_{i=1}^{N}x_i ,
\label{eq:var_chg}
\\
y_j = x_j - \frac{(\sqrt{N}\xcm+x_1)} {\sqrt{N} + 1} .
\end{eqnarray}
With these substitutions, the 1D version of the Schr\"odinger equation (cf. Eq.~\eref{mpscho}) can be rewritten as
\begin{equation}
\label{eq:condi}
\rmi \hbar \frac{\partial \Psi}{\partial t} = \left( - \frac{\hbar^2}{2 \Mcm}\frac{\partial^2}{\partial \xcm^2} -\frac{\hbar^2}{2m} \sum_{i = 2}^N \frac{\partial^2}{\partial y_i^2} + V \right) \Psi ,
\end{equation}
with $\Mcm \equiv N m$ and $\Psi \equiv \Psi(\xcm,\vy,t)$ is the many-particle wave function with the new coordinates.
The coordinates $\vy$ in \eref{eq:var_chg} are chosen such that no crossed terms appear in the Laplacian of \eref{eq:condi},
see \ref{app:centerofmass}. Notice that the many-particle Schr\"odinger equation in \eref{eq:condi} is, in general, non separable because of the potential $V$ defined in \eref{appoten}.

Hereafter, we derive the wave equation associated to the conditional wave function for the center of mass~\cite{conditional,Oriols07,norsen} defined as $\psicd(\xcm,t) \equiv \Psi(\xcm,\vy^j(t),t)$ associated to the $j$-experiment. 
By construction, the velocity (and therefore the trajectory) of the center of mass only depends on the spatial derivatives along $\xcm$~\cite{conditional,norsen}.
Therefore, $\xcm^j(t)$ can be equivalently computed from either $\psicd$ or $\Psi$.
Following Ref.~\cite{Oriols07}, the previous \eref{eq:condi} can be written in the conditional form as
 \begin{eqnarray}  \fl
\label{eq:conditional}
\rmi \hbar \frac{\partial \psicd}{\partial t} = - \frac{\hbar^2}{2 \Mcm} \frac{\partial^2 \psicd}{\partial \xcm^2} -\left.\frac{\hbar^2}{2m} \sum_{i = 2}^N \frac{\partial^2 \Psi(\xcm,\vy,t)}{\partial y_i^2} \right|_{\vy^j(t)}
\nonumber \\
- \left.\rmi\hbar \sum_{i = 2}^N v_i^j(t)  \frac{\partial \Psi(\xcm,\vy,t)}{\partial y_i}\right|_{\vy^j(t)}+ \Vcm(\xcm)  \psicd,
\end{eqnarray}
where $\Vcm(\xcm) = N\Vext(\xcm)$. See \ref{app:centerofmass} to see how the term $V$ in the many-particle Schr\"odinger equation \eref{eq:condi} is translated into the term $\Vcm$ in the conditional wave function \eref{eq:conditional}.
By inserting the polar decomposition of the full and conditional wave functions,
$\Psi \equiv R \exp(\rmi S/\hbar)$ and $\psicd \equiv \Rcd \exp(\rmi \Scd/\hbar)$,
into \eref{eq:conditional}, one can then derive a continuity-like equation,
\begin{eqnarray}
0= \frac{\partial \Rcd^2}{\partial t} + \frac {\partial}{\partial \xcm} \left( \Rcd^2 \frac{\partial \Scd}{\partial \xcm} \frac {1} {\Mcm} \right) +J|_{\vy^j(t)} ,
\label{condition4}
\\
J= \hbar \sum_{i=2}^N \left [ \frac {\partial R^2}{\partial y_i} v_i^j(t) - \frac {\partial}{\partial y_i} \left ( \frac{1}{m} r^2 \frac {\partial S}{\partial y_i} \right ) \right ] ,
\label{condition4bis}
\end{eqnarray}
plus a quantum Hamilton--Jacobi-like equation
\begin{eqnarray}
0= \frac{\partial \Scd}{\partial t} + \frac {1} {2 \Mcm} \left(\frac{\partial \Scd}{\partial \xcm} \right)^2+ \Vcm +G|_{\vy=\vy^j(t)} ,
\label{condition5}
\\
G=\Qcm + \sum_{i=2}^N \left ( \frac{1}{2 m} \left ( \frac {\partial S}{\partial y_i} \right)^2 + Q_i - v_i^j(t) \frac {\partial S}{\partial y_i} \right) .
\label{condition5bis}
\end{eqnarray}
They include the definition of the quantum potentials
\begin{eqnarray}
\label{eq:quantum_cm1}
\Qcm = \Qcm(\xcm,\vy,t)=-\frac{\hbar^2}{2 \Mcm R}\frac{\partial^2 R}{\partial \xcm^2 } ,
\\
\label{eq:quantum_cm2}
Q_{i} = Q_{i}(\xcm,\vy,t)=-\frac{\hbar^2}{2mR}\frac{\partial^2 R}{\partial y_i^2 } ,
\end{eqnarray}
and the (non-local) velocity fields 
\begin{eqnarray}
\vcm = \vcm(\xcm,\vy,t) = \frac{1}{\Mcm} \frac{\partial S}{\partial \xcm} ,\label{vtcm} \\
v_i = v_i(\xcm,\vy,t) = \frac{1}{m}\frac{\partial S}{\partial y_i} .
\end{eqnarray}

The behavior of the quantum Hamilton--Jacobi equation \eref{condition5} would be classical if the effect of the ``potential'' $G$ could be ignored.
Therefore, the key point in our demonstration is to show that $G$ in \eref{condition5bis} fulfills
\begin{equation}
\left.\frac{\partial G}{\partial \xcm}\right|_{\vy=\vy^j(t)} = 0,
\label{G0}
\end{equation}
for a quantum state full of identical particles.
The first part of this proof is showing that
\begin{equation} \fl
\left.\frac {\partial}{\partial\xcm} \sum_{i=2}^N \left ( \frac{1}{2 m} \left ( \frac {\partial S}{\partial y_i} \right)^2 - v_i^j(t) \frac {\partial S}{\partial y_i} \right)\right|_{\vy^j(t)}
= \left.\left ( \frac{1}{m} \frac {\partial S}{\partial y_i} \frac {\partial^2 S}{\partial\xcm y_i} - v_i^j(t) \frac {\partial^2 S}{\partial\xcm y_i} \right)\right|_{\vy^j(t)} = 0 ,
\end{equation}
where he have used that ${\partial S}/{\partial y_i}$ depends on $\xcm$, but $v_i^j(t)$ does not.
The second part of the proof is showing that
\begin{equation}
\left[ \frac {\partial}{\partial \xcm} \left(\Qcm+\sum_{i=2}^N Q_i \right)\right]_{\vy^j(t)} = 0 .
\label{extra3}
\end{equation}
Up to here all equations involve only the $j$-experiment. Since we know from \sref{evolution} that any other trajectory of the center of mass associated to the $k$-experiment will satisfy $\xcm^k(t)=\xcm^j(t) \equiv \xcm(t)$, the shape of the potential term in \eref{extra3} for the $j$-experiment must be also equal to that of any other $k$-experiment.
Therefore, we substitute \eref{extra3} by an average over an ensemble of experiments,
\begin{equation} \fl
\left[ \frac {\partial}{\partial \xcm} \left(\Qcm+ \sum_{i=2}^N Q_i \right)\right]_{\vy^j(t)}= 
\frac {1} {M} \sum_{k=1}^M \left[ \frac {\partial}{\partial \xcm} \left( \Qcm+\sum_{i=2}^N Q_i\right)\right]_{\xcm^k(t),\vy^k(t)} .
\label{clasical2}
\end{equation}
Since the trajectories $\xcm^k(t)$ and $\vy^k(t)$ in the r.h.s. are selected according to \eref{QE},
we can substitute the sum in \eref{clasical2} by an integral weighted by $R^2$,
\begin{eqnarray} \fl
\frac{1}{M} \sum_{k=1}^M \left[ \frac {\partial}{\partial \xcm} \left( \Qcm+\sum_{i=2}^N Q_i\right)\right]_{\xcm^k(t),\vy^k(t)}\nonumber\\=
\int\limits_{\xcm}\int\limits_{y_2}\ldots\int\limits_{y_N} R^2 \frac{\partial}{\partial \xcm} \left( \Qcm+\sum_{i=2}^N Q_i \right ) d\xcm dy_2\ldots dy_N .
\label{condition3new}
\end{eqnarray}
For each term $Q_i$ we have that
\begin{equation} \fl
\int\limits_{\xcm} R^2(\xcm,\vy) \frac{\partial Q_i(\xcm,\vy)}{\partial \xcm} d \xcm  =
\frac{\hbar^2}{2m} \left[ \int\limits_{\xcm} \frac{\partial R}{\partial \xcm} \frac{\partial^2 R}{\partial y_i^2}  d\xcm 
-\int\limits_{\xcm} R  \frac{\partial^3 R}{\partial \xcm \partial y_i^2}  d\xcm \right] .
\label{demo32}
\end{equation}
It can be easily seen that these two terms are equal (but with opposite signs) by integrating by parts the first term (assuming that $R$ is zero for $x\rightarrow\pm \infty$).
Therefore \eref{demo32} is equal to 0.
A similar argument can be made to show that the term with $\Qcm$ in \eref{condition3new} is also zero.
The fact that \eref{condition3new} vanishes can be anticipated by knowing that this type of integrals on the whole configuration space also appear (and are zero) in the derivation of Ehrenfest's theorem if the polar form of the wave function is used.

We have just demonstrated that the (conditional) wave equation of a center of mass associated to a quantum state full of identical particles implies \eref{G0}.
In this case, the Hamilton--Jacobi equation in \eref{condition5} has no dependence on $\Rcd$, and only on $\Scd$. Therefore, the velocity of the center of mass,
\begin{equation}
\vcm= \frac{1}{\Mcm} \frac{\partial \Scd}{ \partial \xcm} ,
\end{equation}
and its trajectory can be computed from \eref{condition5} independently of \eref{condition4}.
Moreover, \eref{condition5} ignoring the ``potential'' $G$ is analogous to the (classical) Hamilton--Jacobi equation,
from which one can derive a Schr\"odinger-like equation
\begin{equation}
\rmi \hbar \frac{\partial \psicd}{\partial t}= \left( -\frac{\hbar^2}{2 \Mcm} \frac{\partial^2}{\partial \xcm^2} + \Vcm - \Qcm \right) \psicd .
\label{condgen}
\end{equation}
In the derivation of this wave equation, we have also used \eref{condition4}. The exact shape of the term $J$ in \eref{condition4} is irrelevant for computing the velocity of the center of mass (which only depends on \eref{condition5}), and we have assumed the term $J=0$ to deal with a conditional wave function with norm equal to one.
This equation is also known as the (non-linear) classical Schr\"odinger wave equation~\cite{richardson_nonlinear_2014,Oriols12,nikolic}.
A study of the dynamics associated with this equation can be found in Ref.~\cite{FNL2016}. We emphasize that the correlations among $\xcm$ and the rest of $y_i$ present in \eref{eq:condi} are included through the non-linear term $-\Qcm$ in the conditional equation of motion \eref{condgen}. 

\subsubsection*{Numerical examples}

\begin{figure}
\centerline{\includegraphics[width=0.6\columnwidth]{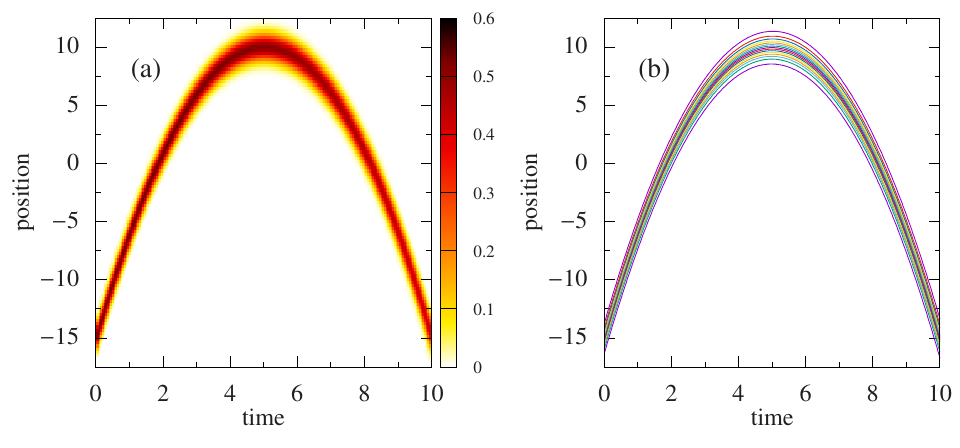}}
\caption{
(a)
Evolution of a classical wave packet subjected to a potential $\Vcm(x)=2 x$.
The initial wave function is a Gaussian wave packet of width $\sigma = 1$, centered around $x_0=-15$, and an initial positive velocity $k_0 = 10$.
(b)
Trajectories corresponding to these dynamics.
Units are $\Mcm=\hbar=1$.
}
\label{fig:C1}
\end{figure}

\begin{figure}
\centerline{\includegraphics[width=0.6\columnwidth]{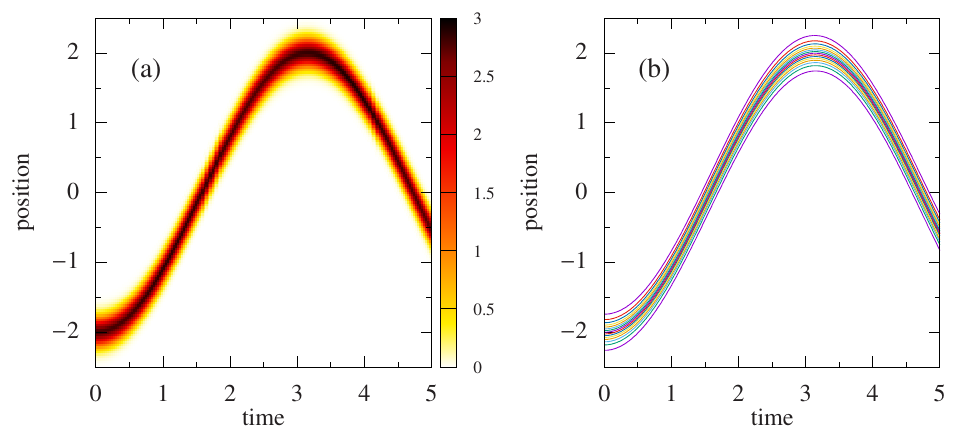}}
\caption{
Same as \fref{fig:C1} but in a potential $\Vcm(x)=x^2/2$.
The initial Gaussian wave function has $x_0=-2$, $\sigma=0.2$, and $k_0=0$.
}
\label{fig:C2}
\end{figure}

In order to illustrate the previous derivation,
in what follows we will solve the (non-linear) classical Schr\"odinger wave equation in \eref{condgen}.
We show in \fref{fig:C1} the case of the evolution of a wave packet under a potential $\Vcm(x) = 2 x$.
One can see that the classical wave packet preserves its shape, and its corresponding trajectories are the expected classical parabolic ones.
This contrasts with the simulation of the same initial quantum wave packet in \fref{fig:Q1}, which expanded over time.
Another simulation is shown in \fref{fig:C2}, in this case for a harmonic potential with an narrow initial wave packet displaced from the origin.
As expected from the classical behavior, the trajectories oscillate around the origin, while the wave packet maintains its narrow shape.
We emphasize that the initial wave packet has to reflect that the probability distribution of the center of mass is very sharp~\cite{FNL2016}.

\section{Quantum states without a classical center of mass}
\label{sec:macroscopic}

There are certainly many examples of quantum states whose center of mass do not behave classically~\cite{Giulini96, schlosshauer14,Oriol11,Zeilinger}.
In the following we discuss two paradigmatic examples.

\subsection{Single-particle states}

For a single particle states, the center of mass in a unique experiment is the Bohmian position of the particle itself.
Moreover, it cannot satisfy condition 1 because different experiments will provide different results.
Therefore, the center of mass of a quantum system with one (or few particles) cannot follow our classical intuition. 

Let us analyze the problems appearing when Bohmian mechanics is used to study the quantum-to-classical transition for a single-particle states.
By inserting $\psi = R \exp(\rmi S/\hbar)$ into the single-particle Schr\"odinger equation one arrives to a quantum continuity equation
\begin{equation}
\frac {\partial R^2} {\partial t} + \frac {\partial } {\partial x} \left( \frac {R^2} {m} \frac{\partial S}{\partial x} \right) =0,
 \label{conti}
\end{equation}
plus a quantum Hamilton--Jacobi equation~\cite{bohm52} given by
\begin{equation}
\label{hamilton}
\frac{\partial S}{\partial t} + \frac {1} {2m} \left(\frac{\partial S}{\partial x} \right)^2 +\Vext + Q = 0.
\end{equation}
It can be easily demonstrated that \eref{conti} and \eref{hamilton} give a Newton-like equation for the (Bohmian) trajectories~\cite{bohm52,Oriols12}
\begin{equation}
\label{newton}
m  \frac {d  v(x^j(t),t)} {dt} = \left[ -\frac {\partial} {\partial x} \left(\Vext + Q \right) \right]_{x = x^j(t)}.
\end{equation}

It has been argued~\cite{Rosen64} that a classical (Newtonian) trajectory could be obtained from \eref{newton} by just adding a \emph{new} condition
\begin{equation}
\frac {\partial Q} {\partial x} = 0.
\label{qzero}
\end{equation}
The problem with this statement is that the classical state given by $x^j(t)$ is not compatible with a quantum state given by the same trajectory $x^j(t)$ and a wave function $\psi$.
The reason of such incompatibility is that $\psi$ does not exist in general. The wave function $\psi$ would have to satisfy,  in each position, three equations, Eqs.~\eref{conti}, \eref{hamilton} and \eref{qzero}, but with only two unknowns, $R$ and $S$.

Another single-particle approach to reach classical dynamics is to interpret the potential $\Vext$ as an additional unknown that allows to define some (exotic) systems where the trajectory and the wave function belong to a state which is simultaneously classical and quantum~\cite{Mako02}.
The simplest example is a plane wave with a constant $R=1$, giving $Q=0$.
However, even these particular compatible solutions have some unphysical features in disagreement with our classical intuition.  The initial position of the Bohmian trajectories $x^j(t)$ associated to these systems obviously have to be selected according to the distribution $|\psi|^2$ obtained from \eref{QE}. This means that different initial positions are obtained in different experiments.
For the plane wave, the particle can depart from anywhere at the initial time, contradicting our classical intuition of having well defined initial positions. 

On the contrary, we have shown in \sref{sec:fullofparticles} that a quantum state full of identical particles is compatible with a center of mass following a classical trajectory.
The reason why both classical and quantum states are compatible in our case is because the condition in \eref{G0} is satisfied in a \emph{natural} way by a quantum state full of identical particles (without imposing any condition on $\Vext$).
In addition, the classical trajectory of the center of mass of such states directly implies that its initial position and velocity do not change when the experiment is repeated. 

\subsection{Many-particle states}
\label{exotic}

Our definition of  a quantum state full of identical particles discussed in \sref{example2} is quite \emph{natural} when the number of particles tends to be very large. However, we define here a quantum state with a large number $N$ of particles with strong correlations that do not satisfy our requirements for a quantum state full of identical particles.

One can think of wave functions of identical particles which make it impossible for a unique experiment to fill the whole support of the marginal distribution.
Macroscopic quantum many-particle superpositions~\cite{Oriol11,Zeilinger,Oriol11a} will not satisfy the condition in \eref{mar1} and therefore we do not expect a classical behavior for their center of mass, even when $N\to\infty$.
An extreme example would be the superposition of two separated wave packets (a Schr\"odinger-cat-like state) such as
\begin{equation}
\Psi(x_1,\ldots,x_N) = \frac{1}{\sqrt{2}} \left(\prod_{i=1}^N \phi(x_i-x_L) +\prod_{i=1}^N  \phi(x_i-x_R)  \right).
 \label{mar17}
\end{equation}
We assume that $\phi(x)$ is a (properly normalized) wave packet centered around $x=0$, whose support is much smaller than the distance between the two wave packets ($x_R-x_L$) so that the overlap between $\phi(x_i-x_L)$  and $\phi(x_i-x_R)$ is zero.
The wave function in \eref{mar17} only allows for two kinds of quantum states.
The first one corresponds to the wave function above plus all particles around $x_L$.
The second one corresponds to the same wave function plus all particles around $x_R$.

In order to see this from the point of view of the probability distributions, we calculate the marginal probability distribution of this state, using \eref{margdef1},
\begin{equation}
D(x,0)=\frac{1}{2} \left(|\phi(x_i-x_L)|^2+|\phi(x_i-x_R)|^2\right).
\end{equation} 
Therefore, the first particle position in the $j$-experiment has equal probability to be in either $x^j_1(0) \approx x_L$ or $x^j_1(0) \approx x_R$.
If for instance it is $x^j_1(0) \approx x_L$, then, using \eref{marcon1} and \eref{marcon2},
the second particle is selected according to $D^{j,2}(x,0)=|\phi(x_i-x_L)|^2$, and it will also be $x^j_2(0) \approx x_L$.
In fact, all subsequent particles are located around $x_L$ because \eref{marcon1} and \eref{marcon2} show that $D^{j,i}(x_i,0)=|\phi(x_i-x_L)|^2$ for $i>1$.
Similarly, if in another experiment the first particle is $x^j_1(0) \approx x_R$, then, all particles will be around $x^j_i(0)\approx x_R$.
It is obvious then, that in this case $D(x,0)\neq D^{j,i}(x,0)$ in all experiments. This is because the marginal distribution for this state has a non-zero support around both $x_L$ and $x_R$, while the quantum state in any experiment involves only particles at left or only particles at the right, but never particles at both sides.

We discuss here why the center of mass of a quantum state like the one in \eref{mar17} can show quantum interference.
Although the marginal distribution has support in both sides, in a particular experiment, the Bohmian trajectories associated to this state will be present in only one side, say the left support.
Thus, the dynamics of the center of mass is associated only to the particles in the left support of the wave function.
However, (classically unexpected) interferences could appear later if the left wave function overlaps and interferes with the right one (empty of particles), thus modifying the velocities of the particles.
On the contrary, in the numerical example of \sref{examfop} where the marginal distribution also has two separated supports, such (classically unexpected) interferences will not appear because it is a quantum state full of identical particles.
Bohmian trajectories will always fill up both left and right supports and the center of mass will always be an average over all (left and right) particles.
If the left and right support are large enough to be macroscopically distinguishable, we will \emph{see} two classical particles, described by the center of mass of the left and right Bohmian particles, respectively.
The trajectories of these centers of mass will correspond to the elastic collision between classical particles.
We conclude that quantum states whose supports are partially empty of particles are required to observe effects against our classical intuition.

\section{Conclusions}
\label{sec:conc}

In summary, by using the peculiar properties of the center of mass interpreted as a Bohmian particle, we have provided a \emph{natural} route to explain the quantum-to-classical transition. We have defined a quantum states full of identical particles as the state whose distribution of the Bohmian positions in a unique experiment is always equal to the marginal distribution.
The center of mass of such states satisfies our classical intuition in the sense that, first, its initial position and velocity are perfectly fixed when experiments are repeated (prepared with the same wave function) and, second, it follows a classical trajectory.
We emphasize that only the center of mass behaves classically, while the rest of microscopic degrees of freedom can and will show quantum dynamics. 
In this sense, the quantum-to-classical transition appears due to the \emph{natural} coarse-graining description of the center of mass.

Due to the compatibility between Bohmian and orthodox results~\cite{bohm52,Holland93,Oriols12,ABM_review}, the arguments in this paper can be equivalently derived using with orthodox arguments.
The Bohmian route explored here avoids dealing with the reduced density matrix and the collapse law.
There is a commonly accepted wisdom in the orthodox attempts that decoherence plays a relevant role in the quantum-to-classical transition, and this work does not contradict this.
One can see that the center of mass (our open system) is strongly entangled with the rest of degrees of freedom of the macroscopic object (the environment). Notice, from the definition of the potential in \eref{appoten}, that the many-particle Schr\"odinger equation in \eref{eq:condi} is, in general, non separable.
Without this entanglement, we will not arrive to the classical (dispersionless) wave equation in \sref{wavequation}, but to a single-particle Schr\"odinger equation with the typical spreading of wave packets. Notice that the original Schr\"odinger equation is linear, while the classical version is non-linear, breaking the superposition principle.  
A paradigmatic example of the role of decoherence in destroying superposition (and avoiding wave packet spreading) was initially presented by Zurek using the example of Hyperion, a chaotically tumbling moon of Saturn~\cite{decoh1,decoh2,decoh3,decoh4}.
He estimated that, without decoherence, within 20 years the quantum state of Hyperion would evolve into a highly nonlocal coherent superposition of macroscopically distinguishable orientations. 
It is important to emphasize that, in our work, the environment of the center of mass of Hyperion would consist of $N \approx 10^{44}$ particles, which would be responsible for the decoherence of the center of mass.

The conclusions in this paper for a quantum state full of identical particles, derived for an infinite number of particles, can be translated into a macroscopic system with a very large but finite number of particles when the error defined in \ref{app:error} remains smaller than some predetermined measuring accuracy.
In particular, for the two numerical examples of this paper, the central limit theorem~\cite{central} ensures that the center of mass of a quantum state full of identical particles with a finite number of particles tends to the exact classical value as $N$ grows.

Finally, an explanation on why we have ignored the measurement apparatus along this article is in order.
It is well-known that the Bohmian formalism does not include any collapse law but, instead one has to include the interaction between the system and a measuring apparatus.
We have ignored this interaction because we are only dealing with a classical object measured by a classical apparatus. 
Both the classical object and the classical measuring apparatus are in a quantum state full of identical particles whose centers of mass follow a classical trajectory $\xcms(t)$ and $\xcma(t)$, respectively.
Then, the interaction between the system and the apparatus, i.e. between $\xcms(t)$ and $\xcma(t)$, is unproblematic and it can be ignored if the type of classical measurement is assumed to not perturb the classical macroscopic object. On the contrary, the present work cannot be directly applied to the measurement of a quantum system in general. Obviously, many quantum systems cannot be described by a quantum state full of identical particles when different experiments (with identical wave function preparation) provide different measured results. Nevertheless, a straightforward generalization of the present work can explain why the measuring apparatus (entangled with the quantum system) presents a classical behavior with its macroscopic pointer (in fact, its center of mass) following a classical trajectory. 

\ack
We would like to thank David Tena for fruitful discussions.
This work was supported by This work has been partially supported by the Fondo Europeo de Desarrollo Regional (FEDER) and the ``Ministerio de Ciencia e Innovaci\'{o}n'' through the Spanish Project TEC2015-67462-C2-1-R,
the Generalitat de Catalunya (2014 SGR-384),
the European Union's Horizon 2020 research and innovation program under grant agreement No 696656,
and the Okinawa Institute of Science and Technology Graduate University.

\appendix

\section{Evolution of the error of the center of mass for a quantum state full of identical particles with a finite number of particles.}
\label{app:error}

A definition of a quantum state full of identical particles in \eref{mar1} of the text, in principle, requires  $N \to \infty$. Let us now study the properties of a quantum state with a finite number, $N_F$, pf particles that becomes a quantum state full of identical particles when $N_F \to \infty$. We use the subscript $F$ in $N_F$ to remind that the number of particles is finite.
In particular, the selection of the initial position of the trajectories associated of these new quantum state with only $N_F$ particles follows also \eref{marcon1} and \eref{marcon2}.
Once the $N_F$ particles are selected, we can distribute them following
\begin{equation}
C^{j_0,F}(\vec r,t) =  \frac{1}{N_F} \sum_{i=1}^{N_F} \delta(\vec r-\vec r_i^{j_0}(t)) ,
\label{disF}
\end{equation}
and define their center of mass as
\begin{equation}
\rcm^{j_0,F}(t)
= \int d\vec r \; \vec r\;  C^{j_0,F}(\vec r,t)=\frac{1}{N_F}\sum_{i=1}^{N_F} \vec r_i^{j_0}(t) .
\label{cmf}
\end{equation}
Notice again that $\rcm^{j_0,F}(t)\neq \rcm^{j_0}(t)$ because we are dealing here with a finite number of particles $N_F$, while we know that $\rcm^{j_0}(t)=\langle \rcm \rangle(t)$. The error resulting from comparing this center of mass $\rcm^{j_0,F}(t)$ with the one obtained for $N_F\to\infty$, can be estimated as
\begin{equation}
\Err(t) = \left|\langle \rcm \rangle (t)-\rcm^{j_0,F} (t)\right| .
\label{error} 
\end{equation}
As indicated in \eref{qsfip},
$\langle \rcm\rangle (t)$
is independent of the experiment, but $\rcm^{j_0,F} (t)$ in \eref{cmf} varies between experiments due to quantum randomness.

To further develop expression \eref{error}, let us assume now that the selections of all $\vec r_i^{j_0}(t)$ are independent, i.e.,
we select each $\vec r_i^{j_0}(t)$ according to $D(\vec r_i,t)$.
This is exactly the case in the two numerical examples explained in sections \ref{example1} and \ref{examfop}. The center of mass in \eref{cmf} corresponds to a sequence of independent and identically distributed random variables $\vec r_i$ drawn from a distribution  $D(\vec r_i,t)$ with a mean value given by $\langle \rcm\rangle (t)=\int \vec r\;  D(\vec r,t) d\vec r$ and with a finite variance given by
\begin{equation}
\sigma^2(t)= \int (\vec r-\langle \rcm\rangle (t))^2\;  D(\vec r,t) d\vec r  .
\end{equation}
We know from the central limit theorem~\cite{central} that the distribution of $\rcm^{j,F} (t)$ in different experiments given by \eref{cmf} follows a normal distribution when $N_F$ grows with mean value and variance
\begin{eqnarray}
\rcm^{j_0,F}(t)= \int d\vec r \; \vec r\;  C^{j_0,F}(\vec r,t) \approx \langle \rcm\rangle (t)  , \\
\int d\vec r \; (\vec r-\langle \rcm\rangle (t))^2\;  C^{j_0,F}(\vec r,t) \approx \frac{\sigma(t)^2}{N_F} .
\label{limvar}
\end{eqnarray} 
These results are valid for any initial distribution $D(\vec r_i,t)$ as far as $N_F$ is large enough.\\

The error in expression \eref{error} can now be rewritten in terms of the probability of getting a difference between $\langle \rcm \rangle (t)$ and $\rcm^{j_0,F} (t)$ smaller than a given error, $\Err$,
\begin{equation}
{\mathcal P}\left(\left|\langle \rcm \rangle-\rcm^{j,F}\right|<\Err\right)
=2 \Phi_N\left(\frac{\sqrt{N_F} \Err}{\sigma}\right)-1
\label{error1}
\end{equation}
where $\Phi_N(x)$ is the cumulative distribution function of the standard normal distribution,
\begin{equation}
\Phi_N(x)=\int_{-\infty}^{x} \frac{1}{\sqrt{2\pi}} \exp(-t^2/2) dt ,
\end{equation}
and we have used its property $\Phi_N(x)+\Phi_N(-x)=1$.
If we require, for example, the difference $\langle \rcm \rangle-\rcm^{j_0,F}$ be smaller than $\Err=0.005\sigma$ with a probability of ${\mathcal P}\left(\left|\langle \rcm \rangle-\rcm^{j_0,F}\right|<0.005\sigma\right)=0.98$, then, we get that the number of particles $N_F$ has to be equal or larger than:
\begin{equation}
N_F \ge \frac{\left(\Phi_N^{-1}(0.99)\right)^2}{0.005^2}
\simeq 2 \times 10^5
\label{error2}
\end{equation}
In summary, if we consider $0.005\sigma$ an acceptable error for $\rcm^{j,F}$, then we are sure than $98\%$ of the experiments with our quantum state with a number of particles $N_F\gtrsim 2\times 10^5$ satisfy the fixed error.

As a more realistic example, let us consider a macroscopic system with the number of particles equal to a mol of the matter, i.e. $N_F=6\times10^{23}$ particles.
In addition, we require that the value of $\rcm^{j,F}$ gives always the classical value, i.e., 
that only once in $M_F = 2\times10^{12}$ experiments, the value of $\rcm^{j,F}$ overcomes a fixed value of the  $\Err$. Then, we can compute the required error by solving the relation
${\mathcal P}=1 - 1/M_F$
in \eref{error1} as:
\begin{equation}
\frac{\Err}{\sigma}
=\frac{\Phi_N^{-1}(1-10^{-12})}{\sqrt{N_F}}
\simeq 9\times10^{-12}  
\label{error3}
\end{equation}
In summary, for a quantum state with a number of particles typical of a macroscopic system, i.e. $N_F=6\times10^{23}$, the error of $\rcm^{j,F}$ is smaller than $\Err \approx10^{-11} \sigma$ always (except  in one experiment every $M_F=2\times10^{12}$). 

The time evolution of the error in \eref{error} can be obtained once we know the particular time-dependence of the variance of $D(x,t)$. For example, in the case of $D(x,t)$ given by the modulus square of a Gaussian wave packet in free space, then,  the standard deviation is given (for larger times) by
\begin{equation}
\sigma(t)
= \sigma_0 \sqrt{1 + \left( \frac{\hbar t}{2 m \sigma_0^2} \right)^2} \approx  \frac{\hbar t}{2 m \sigma_0} .
\label{sigma}
\end{equation}
For example, assuming an initial spatial dispersion $\sigma_0=100$ nm, a mol of carbon atoms ($m=2\times10^{-26}$ kg), after $t = 1$ year of classical evolution, the absolute error in \eref{error3} is given by $\Err (t) \simeq 10^{-12} \sigma(t) \simeq 8$ $\mu$m.
In summary, in the overwhelming majority of experiments (all $M_F=2\times10^{12}$ experiments except one), the error in the center of mass after one year of evolution, between the exact value (with $N\to\infty$) and the approximate center of mass (with $N_F=6\times10^{23}$) for the described quantum state is smaller than 10 $\mu$m. 

Certainly, in this example $\Err(t)$ grows with time due to the intrinsic expansion of a free wave packet.
However, we want to emphasize that our classical intuition is based on crystalline materials where particles have an ordered structure due to their attractive interactions. Thus, classical objects (i.e. its particles) will tend to remain much more localized than in the above example.
These interactions will also introduce correlations among the different particles and, in principle, the assumption that the selection of all $\vec r_i^{j_0}(t)$ are independent might not seem fully rigorous.
However, one can argue that in a realistic classical system, with $N_F\simeq6\times10^{23}$ interacting particles, the accurate selection of a the first, say $N_F/100$, particles with the  procedure in \eref{marcon1} and \eref{marcon2} will be roughly independent.
This is due to the selection of points in a huge (and basically empty) configuration space of $3 N_F \sim 10^{24}$ dimensions.
Only the selection of the last particles will be influenced by the non-negligible correlations with the previous ones.

\section{Wave equation for the center of mass coordinates}
\label{app:centerofmass}

Our aim here is to find a change of coordinates in the 1D many-particle Schr\"odinger equation, cf. 1D version of Eq.~\eref{mpscho}, with the usual definition of the center of mass,
\begin{equation}
\label{eq:ap_chg1}
\xcm = \frac{1}{N} \sum_{i=1}^N x_i .
\end{equation}
and without cross terms appearing in the Laplacian.
The additional set of $N-1$ coordinates can be written as
\begin{equation}
\label{eq:ap_chg2}
y_j = \sum_{i=1}^N \alpha^{(j)}_{i} x_i \qquad \textrm{for} \ j=2,\ldots,N ,
\end{equation}
and the $\alpha^{(j)}_{i}$ will be fixed by the condition that cross terms do not appear in the Laplacian
\begin{equation}
\sum_{i=1}^N \frac{\partial^2 \psi}{\partial x_i^2} = \frac{1}{N} \frac{\partial^2 \psi}{\partial \xcm^2} +\sum_{j=2}^N \frac{\partial^2 \psi}{\partial y_j^2} .
\label{eq:laplacian}
\end{equation}

Substituting Eqs.~\eref{eq:ap_chg1} and \eref{eq:ap_chg2} into the l.h.s. of \eref{eq:laplacian}, one obtains
\begin{equation} \fl
\sum_{i=1}^N \frac{\partial^2 \psi}{\partial x_i^2} =
\frac{1}{N} \frac{\partial^2 \psi}{\partial \xcm^2}
+ \frac{2}{N} \sum_{k=2}^N \left[  \frac{\partial^2 \psi}{\partial \xcm \partial y_k} \sum_{i=1}^N \alpha^{(k)}_{i} \right]
+ \sum_{k=2}^N \sum_{j=2}^N \left[ \frac{\partial^2 \psi}{\partial y_j \partial y_k} \sum_{i=1}^N \alpha^{(j)}_{i} \alpha^{(k)}_{i} \right] .
\label{eq:whatwehave}
\end{equation}
Comparing this with \eref{eq:laplacian} we see that
the conditions for our change of variables are
\begin{equation} \fl
\label{eq:base_conds}
0 = \sum_{i=1}^N \alpha^{(j)}_{i} , \quad
1 = \sum_{i=1}^N \Big(\alpha^{(j)}_{i}\Big)^2 , \quad
0 = \sum_{i=1}^N \alpha^{(j)}_{i} \alpha^{(k)}_{i} \ \textrm{for} \ j \neq k .
\end{equation}

We propose a change of variables with the following structure (using $x_1$ separately as we only need $N-1$ variables besides the center of mass):
\begin{equation} \fl
y_j = a x_j + b \xcm + c x_1 = a x_j + \frac{b}{N} \sum_{i=1}^N x_i + c x_1
\quad \Rightarrow \quad 
\alpha^{(j)}_{k} = a \, \delta_{jk}  + \frac{b}{N} + c \, \delta_{1k}
\end{equation}

We impose conditions \eref{eq:base_conds} in order to get the following system
\begin{eqnarray} 
\fl
0 = \sum_{i=1}^N \alpha^{(j)}_{i}
= a + b + c ,
\\
\fl
1 = \sum_{i=1}^N \Big(\alpha^{(j)}_{i}\Big)^2
= \left(c + \frac{b}{N}\right)^2 + \left(a + \frac{b}{N}\right)^2 + (N-2) \left(\frac{b}{N}\right)^2 ,
\\
\fl
0 = \sum_{i=1}^N \alpha^{(j)}_{i} \alpha^{(k)}_{i}
= \left(c + \frac{b}{N}\right)^2 + \frac{2 b}{N} \left(a + \frac{b}{N}\right) + (N-3) \left(\frac{b}{N}\right)^2 .
\end{eqnarray}
This can be solved to yield the variable changes in Eq.~\eref{eq:var_chg} and the final many-particle Schr\"odinger equation in \eref{eq:condi}.

Now, in order to see how the term $V$ in the many-particle Schr\"odinger equation \eref{eq:condi} is translated into the term $\Vcm$ in the conditional wave function \eref{eq:conditional}, we invert \eref{eq:var_chg} to obtain
\begin{equation}
x_1 = \xcm-\frac{1}{\sqrt{N}} \sum_{i=2}^{N} y_i ,
\label{eq:var_chg_inv}
\qquad
x_j = \xcm +y_j - \frac{1} {\sqrt{N} +N} \sum_{i=2}^{N} y_i.
\end{equation}

We can now rewrite the potential \eref{potential} as: 
\begin{eqnarray}
\fl
V(\xcm, \vy)
=
\Vext\left(\xcm-\frac{1}{\sqrt{N}} \sum_{i=2}^{N} y_i\right)
+ \sum_{j=2}^{N} \Vext\left(\xcm +y_j - \frac{1} {\sqrt{N} +N} \sum_{i=2}^{N} y_i \right)\\
+ \frac{1}{2}\sum_{j=2}^{N}\Vint\left(-\frac{1}{1+\sqrt{N}} \sum_{i=2}^{N} y_i-y_j \right)
+ \frac{1}{2}\sum_{i=2}^N\sum_{{j=2; i\neq j}}^{N}\Vint(y_i-y_j)\nonumber
\label{appoten}
\end{eqnarray}
The terms $\Vint$ have no dependence on $\xcm$.
Therefore, when considering the conditional wave function of the center of mass with $\vec y=\vec y(t)$ in \eref{appoten}, they will just become a purely time-dependent potential.
Their only effect will then be a pure time-dependent phase in the wave function, which can be neglected in the computation of the conditional equation of motion of the center of mass.

Each of the other two terms $\Vext$ in \eref{appoten} have a dependence on $\xcm$ plus a dependence on $\sum_{i=2}^{N} y_i$. We provide a Taylor expansion around $\xcm$ 
\begin{equation} \fl
\Vext(\xcm+\Delta x) = \Vext(\xcm) + \left.\frac{\partial \Vext(x)}{\partial x}\right|_{x=\xcm} \Delta x + \left. \frac{1}{2} \frac{\partial^2 \Vext(x)}{\partial x^2}\right|_{x=\xcm} \Delta x^2 +\ldots .
\label{taylor}
\end{equation}
We define, in order to simplify, the expressions,
\begin{equation}
\beta(\xcm) = \left.\frac{\partial \Vext(x)}{\partial x}\right|_{x=\xcm} ,
\qquad
\gamma(\xcm) = \frac{1}{2} \left.\frac{\partial^2 \Vext(x)}{\partial x^2}\right|_{x=\xcm} .
\end{equation}
This allows to rewrite the part of the potential that depends on $\xcm$ as
\begin{eqnarray} \fl
\Vext\left(\xcm-\frac{1}{\sqrt{N}} \sum_{i=2}^{N} y_i\right)
+ \sum_{j=2}^{N} \Vext\left(\xcm +y_j - \frac{1} {\sqrt{N} +N} \sum_{i=2}^{N} y_i \right)\\
=
N \Vext(\xcm)
+ \beta(\xcm) \left( 1 - \frac{1}{\sqrt{N}} - \frac{N-1} {\sqrt{N} +N} \right) \sum_{i=2}^{N} y_i \nonumber \\
+ \gamma(\xcm) \left[
\sum_{j=2}^{N} y_j^2
+ \left( \frac{1}{N} 
+ \frac{N-1} {(\sqrt{N} +N)^2} 
- \frac{2} {\sqrt{N} +N} \right) \left(\sum_{j=2}^{N} y_j \right)^2
\right] +\ldots \nonumber
\end{eqnarray}

We see that the factor of $\beta(\xcm)$ is zero, i.e. 
$1- \frac{1}{\sqrt{N}} - \frac{N-1} {\sqrt{N} +N} = 0$,
and the factor of $\gamma(\xcm)$ can be simplified as
$\frac{1}{N} + \frac{N-1} {(\sqrt{N} +N)^2} - \frac{2} {\sqrt{N} +N} = 0$,
so we arrive at
\begin{eqnarray} \fl
\Vext\left(\xcm-\frac{1}{\sqrt{N}} \sum_{i=2}^{N} y_i\right)
+ \sum_{j=2}^{N} \Vext\left(\xcm +y_j - \frac{1} {\sqrt{N} +N} \sum_{i=2}^{N} y_i \right)=\nonumber\\
=N \Vext(\xcm) + \gamma(\xcm) \sum_{i=2}^{N} y_i^2 + \ldots
\end{eqnarray}
The $\gamma(\xcm)$ in the second term and higher orders still have, in principle, some $\xcm$ spatial dependence.
We invoke now condition 2 (see \sref{evolution}) that assumes a quadratic approximation for the (long range) external potential, with a negligible dependence of $\gamma$ on $\xcm$.
This means that  $\gamma(\xcm)=\gamma$
and the rest of higher order derivatives of the Taylor expansion become zero.
Under such conditions, when calculating the conditional wave function of the center of mass at $\vec y(t)$, the term $\gamma \sum_{i=2}^{N}y_i^2(t)$ can be neglected as a purely time-dependent term (as happened for the previously discussed $\Vint$ terms).
Therefore, we finally get the external potential of the equation of motion of the conditional wave function of the center of mass 
\begin{equation}
\sum_{j=1}^N\Vext(x_j) \Big|_{\vec y=\vec y(t)} = N\; \Vext(\xcm) \equiv \Vcm(\xcm) .
\label{appoten8}
\end{equation}

The same simple potential can be exactly recovered for a quadratic external potential $\Vext(x)=\alpha+\beta x+ \gamma x^2$ with constant $\alpha$, $\beta$,  and $\gamma$.
Notice that our derivation above demands a more relaxed condition on $\Vext$, as it only requires that this shape (constant $\gamma$) happens along the extension of the object in physical space.

\end{document}